# d-like Symmetry of the Order Parameter and Intrinsic Josephson Effects in Bi2212 Cross-Whisker Junctions

Yoshihiko Takano, [a,b,c] Takeshi Hatano, [a,b,c] Akihiro Fukuyo, [a,d] Akira Ishii, [a,b] Masashi Ohmori, [a,d]
Shunichi Arisawa, [a,b,c] Kazumasa Togano, [a,b,c] and Masashi Tachiki [a,b,c]

[a] *National Institute for Materials Science, Sengen, Tsukuba 305-0047, Japan*
[b] *National Research Institute for Metals, Sengen, Tsukuba 305-0047, Japan*
[c] *CREST, Japan Science and Technology Corporation, 2-1-6, Sengen, Tsukuba 305-0047, Japan*
[d] *Faculty of Science and Technology, Science Univirsity of Tokyo, Yamazaki, Noda 278-8510, Japan*

An intrinsic tunnel junction was made using two Bi-2212 single crystal whiskers. The two whiskers with a cross-angle were overlaid at their *c*-planes and connected by annealing. The angular dependence of the critical current density along the *c*-axis is of the d-wave symmetry. However, the angular dependence is much stronger than that of the conventional d-wave. Furthermore, the current vs. voltage characteristics of the cross-whiskers junctions show a multiple-branch structure at any cross-angle, indicating the formation of the intrinsic Josephson junction array.

Much attention has paid to the symmetry of the order parameter for the high-$Tc$ superconductors. It is necessary to clarify the symmetry of the order parameter to elucidate the mechanism of high-$Tc$ superconductors. In recent years, the results of many experiments have made it generally acceptable to assume that high-$Tc$ superconductors are of a d-wave symmetry [1]; however, the idea that the order parameter is of s-wave still remains tenaciously.

Josephson effects using twisted tunnel junctions directly reflect the pairing symmetry of the order parameters in superconductors. If two superconductors were joined at various angles at their *c*-planes, the Josephson current would be changed according to the symmetry of the order parameters between the two superconductors. In the YBa$_2$Cu$_3$O$_{7-\delta}$ *ab*-plane grain boundary junction formed by the bi-crystal substrate, the angular dependence of the critical current density was observed. The critical current density was exponentially suppressed with the grain boundary angle. This result suggests that the order parameter has a d-wave pairing symmetry [2-6].

On the other hand, Li et al. [7] fabricated twist junctions, in which cleaved single crystal strips were connected by heating with a rotation angle around the *c*-axis. Their junction showed angular independent critical current densities. From their results, they concluded that the pairing is of an s-wave symmetry. In their study, the widths of their single crystal strips were about 300μm. Therefore, the current may possibly to flow on the surface of the sample, since their junction area is much larger than the penetration depth. They heated their samples just below the melting temperature for 30hours. Annealing at such a high temperature for a long time would result in a welding rather than a joining.

We insist that the angular dependence of critical current density ($Jc$) must be examined with the junctions that show intrinsic Josephson effects, since the junction showing intrinsic Josephson properties suggests an almost homogeneous flow of the Josephson current and thus a good quality of the joining [8-10]. Therefore, we performed an intrinsic junction experiment by using Bi2212 whiskers. Two whiskers were joined crosswise and annealed for a short time (30min) at 850ºC, which is 30 degrees lower than the melting point. Because the widths of the whiskers ranged from 10 to 30 μm, the junction area was two orders of magnitude smaller than that of their twist junctions [7]. The multiple-branch structures observed in the current-voltage (I-V) characteristic of the cross-whiskers junction are a feature of the intrinsic Josephson effect [11,12]. In this letter, a d-like pairing symmetry of the critical current density along with the intrinsic Josephson properties is presented using cross-whiskers junctions.

The fabrication process of a cross-whiskers junction is explained in the following. First of all, whisker crystals were prepared. Details of growth condition and their characterization have been given in our previous papers [13-17]. An amorphous plate of $Bi_3Sr_2Ca_2Cu_4O_x$ prepared by the melt quench process was annealed at 850°C for 120h in a flowing 72%$O_2$-$N_2$ gas mixture. Many pieces of whisker crystals were grown from the surface of the amorphous plate. The whisker is single crystal, and the obtained whisker is flat, long, and slender. The direction of length is the *a*-axis, and the flat surface is the *c*-plane. The thickness is 1-3 µm, and the width is 10-30 µm. The length of the large whisker is 20 mm or more.

A schematic of the fabrication process of cross-whiskers junction is shown in Fig. 1. Two appropriate pieces of whiskers were intersected and placed flat on the MgO substrate. The two whiskers were then mutually contacted by the *c*-plane. To study the angular dependence of intrinsic Josephson properties, samples with various cross-angles were made. Suitable cross-angles were chosen in the range of α=45-90°. Here, α represents the interior angle between the two crossing whiskers. To join two whiskers, a short time annealing was performed at 850°C for 30 min in a flowing 70%$O_2$-Ar gas mixture. The cross-section of the junction was polished with a focused ion-beam (FIB) milling machine using Ga. The cross-section was observed with a scanning ion-beam microscope (SIM) in order to examine the conditions of the joint of the two whiskers. The top view of the junction area is shown in Fig. 2(a). The enlargement of side view of the cross-section is shown in Fig. 2(b). No crack or boundary line was visible in the cross-section between upper and lower whiskers. This suggests that annealing process satisfactorily combines the upper and lower whiskers.

The exposed surface acted as an insulating, which is likely due to an insufficiency of Bi because of vaporization during the heat treatment. The stacking for the intrinsic Josephson is constructed of several layers near the intersection of two whiskers, as shown in Fig. 3. This region is electrically isolated in the *a* and *b* directions by insulating surfaces, and, therefore, a current flows along the *c*-axis in the junction as plotted in this figure.

Gold lead wires were attached to the four ends of the cross-whiskers by silver paste for electronic measurements. Here, the whiskers work to form a junction and serve as electronic leads to the junction. The transport properties of the inter-whisker junction were measured by the sophisticated four-probe method. The I-V characteristics were measured in a current-biased mode.

The I-V characteristics of the cross-whiskers junctions with the cross-angle of α=~90° measured at 5K are shown in Fig. 4(a). The multiple-branch structure that is a feature of the intrinsic Josephson junction was observed. The magnitude of the first voltage jump is ~15 mV, corresponding to the typical Bi2212 intrinsic Josephson junction [18]. The interval of the branch became smaller in the high voltage region [18]. The critical current at zero-voltage is approximately 11mA. The area of the junction was measured at 938µm$^2$. From the critical current and the junction area, the critical current density ($J_C$) at 5K was estimated to be ~1170A/cm$^2$. This value is consistent with the typical $J_C$ observed in Bi2212 intrinsic Josephson junctions fabricated on single crystals [8].

Figure 4(b) and (c) display the I-V characteristics measured at 5K for the cross-whiskers junctions with the cross-angles of α=~75° and ~60°, respectively. Clear multiple-branch structures and large hysteresis were also observed. The voltage interval of the branches became smaller with increasing the voltage. The critical current at zero voltage were 341 A/cm$^2$ (α=~75°) and 57 A/cm$^2$ (α=~60°), respectively. The angle dependence of $J_C$ is plotted in Fig. 5. The maximum of $J_C$ occurs at ~90°, and the $J_C$ decreased dramatically with decreasing the cross-angle; the minimum of the $J_C$ appears around 45°. The angular dependence of the $J_C$ has 4-fold symmetry. This fact indicates that the order parameter in the Bi2212 superconductors is of a d-like symmetry.

We assume that the cross-angle dependence of $J_C$ comes from anisotropy of the order parameter and the transfer integral.

Provided that the angular dependence of the transfer integral between the twisted layers is written as $T(\alpha)$, the $Jc$ is expressed as

$$J_C = AT^2(\alpha)\int_0^{2\pi} \Delta_1(\theta)\Delta_2(\theta+\alpha)d\theta \quad (1)$$

Here, α represents the cross-angle between the two whiskers. If the pairing symmetry of the order parameter is of the conventional $d_{x^2-y^2}$ wave, the gap functions $\Delta(\theta)$ for the joined two whiskers take the forms

$$\Delta_1 = \Delta(\cos^2\theta - \sin^2\theta) \quad (2)$$
$$\Delta_2 = \Delta(\cos^2(\theta+\alpha) - \sin^2(\theta+\alpha)) \quad (3)$$

We assume that the momentum in the *a-b* plane for the *c*-axis pair tunneling across the twisted boundary is conserved. Inserting these order parameter into Eq(1) and integrating the equation with respect to θ, the *Jc* can be expressed in terms of the joint angle as follows [1],

$$J_C = \pi AT^2(\alpha)\Delta^2 \cos(2\alpha) \quad (4)$$

When we fix the parameter $AT^2(\alpha)$ by the experimental value and assume that $T(\alpha)$ is independent of $\alpha$, the calculated value of the *Jc* having 4-fold symmetry is plotted by the broken line in Fig. 5. The observed angular dependence is much stronger than the one that calculated. The deviation of the measured angular dependence from the broken line may come from the angular dependence of $T(\alpha)$. If the angular dependence of $T(\alpha)$ is known exactly, we can obtain the structure of the superconducting order parameter which provides important information to elucidate the mechanism of high-$Tc$ superconductivity. The transfer integrals can be measured by the c-axis normal resistivity. In the normal state, all the junctions in the region of the intrinsic Josephson junction, shown with a black thick line in Fig. 3, contribute to the normal resistivity. The contribution from the twisted junction is only 1/N of the resistivity, N being the number of the intrinsic junctions. Since it is difficult to estimate the resistivity from the twisted boundary alone, the determination of $T(\alpha)$ remains to a future problem.

In conclusion, we have successfully fabricated a cross-whiskers junction to make an intrinsic Josephson junction. The junction showing intrinsic Josephson properties suggests the homogeneity of the Josephson current and, thus, a good quality joint. The symmetry of the superconducting order parameter was examined by the cross-angle dependence of a $Jc$ in the cross-whiskers junctions. The $Jc$ was dramatically reduced with decreasing the cross-angle from 90 to 45 degrees. The angular dependence of the $Jc$ is a d-wave like, but much stronger than the angular dependence of the conventional d-wave. If $T(\alpha)$ is known and the angular dependence of the order parameter is determined, these experimental results provide crucially important information concerning the mechanism of the cuprate high-$Tc$ superconductivity.

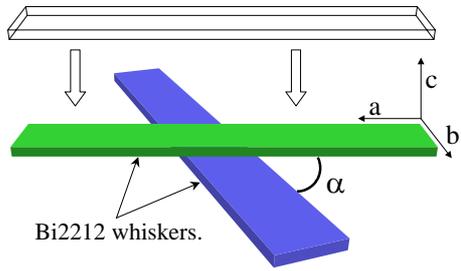

Fig. 1. Schematic of the fabrication process of the cross-whiskers junction. Two whiskers are jointed at various cross-angles.

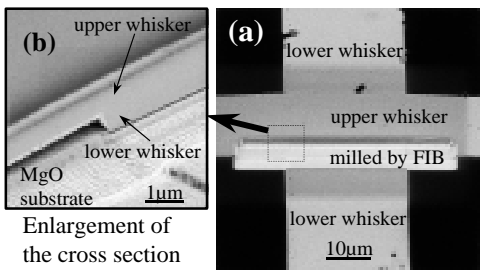

Fig. 2. SIM image of a cross-section of the cross-whiskers junction. The Fig. (b) is the enlargement of the cross-section.

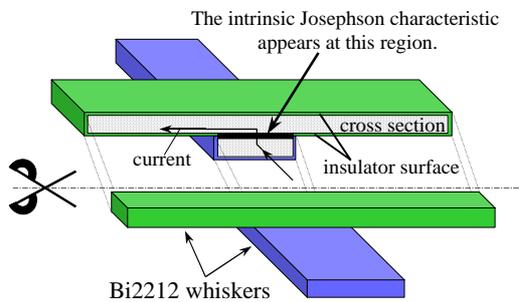

Fig. 3. Schematic of the formation of the mesa structure.

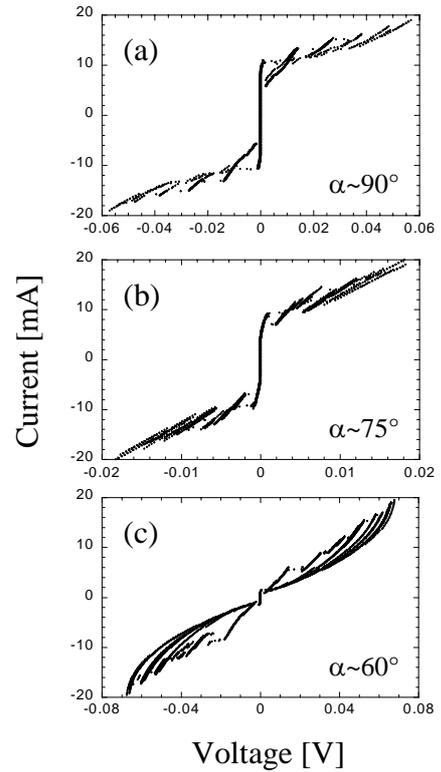

Fig. 4. I-V characteristics of the cross-whiskers junctions with various cross-angles.

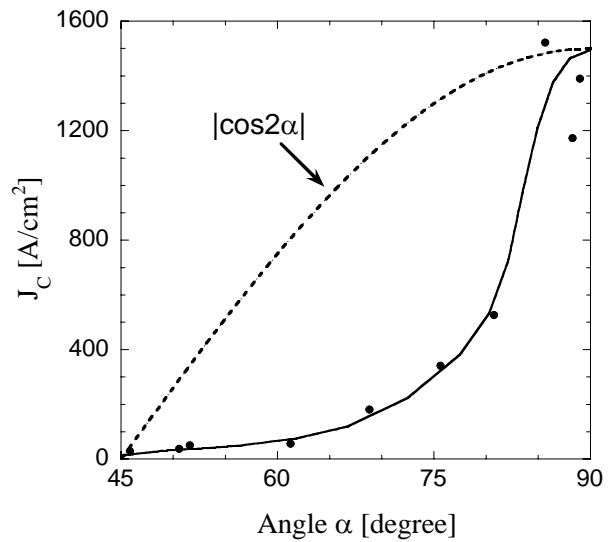

Fig. 5. Angular dependence of the critical current density $J_C$.